\documentstyle[12pt]{article}

\makeatletter
\@addtoreset{equation}{section}
\makeatother

\def\be{\begin{equation}}
\def\ee{\end{equation}}
\def\bea{\begin{eqnarray}}
\def\eea{\end{eqnarray}}

\newcommand{\bee}{\begin{eqnarray}}
\newcommand{\eee}{\end{eqnarray}}
\newcommand{\nn}{\nonumber}

\newcommand{\by}{\bar{y}}

\newcommand{\go}{\omega}
\newcommand{\e}{\epsilon}
\newcommand{\half}{\frac{1}{2}}
\newcommand{\ga}{\alpha}
\newcommand{\gal}{\alpha}

\newcommand{\dga}{{\dot{\alpha}}}
\newcommand{\dgb}{{\dot{\beta}}}
\newcommand{\gb}{\beta}
\newcommand{\gga}{\gamma}

\newcommand{\gl}{\lambda}

\newcommand\un{{\underline{n}}}

\newcommand\um{{\underline{m}}}

\newcommand\ls{\!\!\!\!\!\!\!}

\newcommand {\al} {\alpha}
\newcommand {\dal} {\dot \alpha}
\newcommand {\bt} {\beta}
\newcommand {\dbt} {\dot \beta}

\title{Nt-yet}
\begin{document}

\begin{flushright}
\vspace{1mm}
FIAN/TD/31-99\\
 December { 1999}\\
hep-th/0001031 \\
\end{flushright}

\vspace{1cm}

\begin{center}
{\large\bf
STAR-PRODUCT AND MASSLESS FREE FIELD DYNAMICS IN $AdS_4$}\\
\vspace{3cm}

{\bf K.~I.~Bolotin and M.~A.~Vasiliev}\\
\vspace{2cm}

I.E.Tamm Department of Theoretical Physics, Lebedev Physical Institute,\\
Leninsky prospect 53, 117924, Moscow, Russia
\vspace{4cm}
\end{center}

\begin{abstract}
\baselineskip .4 true cm
\noindent
Generic solution of free equations for massless fields
of an  arbitrary spin  in $AdS_4$
is built in terms of the star-product algebra with spinor
generating elements. A class of ``plane wave"  solutions
is described explicitly.
\end{abstract}

\section {Introduction}

It has been shown (see \cite{rev} for a recent review) that the dynamics of
massless fields of all spins in $AdS_4$ can be  described in terms
of star-product algebras acting on the auxiliary spinor variables.
In this formalism the part of the (nonlinear)
equations of motion that contain
space-time derivatives has a form of zero-curvature or
covariant constancy conditions and
therefore can be solved explicitly at least locally. The aim of
this paper is to illustrate how this machinery can be used in practice
to derive solutions of the massless free field equations in $AdS_4$.

\subsection {AdS Geometry}
As is well-known, $AdS_d$ geometry
is described by the zero-curvature equations for the $AdS_d$ algebra
$o(d-1 ,2)$ with the gauge fields $A_\un{}^{AB}(x)$ ($A,B=0\div d$;
$\um,\un =0\div d-1$; $x^\un $ are coordinates of $AdS_d$)
identified with the $AdS_d$
gravitational fields according to
\be
\omega_\un{}^{a b}=A_\un{}^{a b}\,,\qquad h_\un{}^a=\lambda^{-1} A_\un{}^{a
d}\,,
\ee
where $a,b=0\div d-1 $ and $\lambda$ is a constant to be
identified with the inverse $AdS$ radius. The $o(d-1,2)$
field strengths are
\be
\label{r1}
R_{\um\un}{}^{ab}=\partial_{\um}\go_{\un}{}^{ab}
+\go_{\um}{}^a{}_c\,\go_{\un}{}^{cb}
-\gl^2 h_{\um}{}^a\,h_{\un}{}^b - (\um\leftrightarrow \un) \,,
\ee
\be
\label{r2}
R_{\um\un}{}^a=\partial_{\um}h_{\un}{}^a
+\go_{\um}{}^a{}_c h_{\un}{}^c - (\um\leftrightarrow \un) \,.
\ee
Interpreting $\go_{\un}{}^{ab}$ as Lorentz connection and $h_{\un}{}^b$
as the frame 1-form one observes that $R_{\um\un}{}^a$ identifies with
the torsion tensor while the $\lambda -$independent part of
$R_{\um\un}{}^{ab}$ is the Riemann tensor.
Setting $R_{\um\un}{}^a=0$ one expresses
$\go_{\un}{}^{ab}$ in terms of $h_{\um}{}^a$. Imposing the equation
$R_{\um\un}{}^{ab}=0$ is then equivalent to the equation for $AdS_d$
described in terms of the frame field $h_{\un}{}^a$ which is required to
be non-degenerate. Thus, $AdS_d$ geometry is described by the zero-curvature
equation
\be
R^{AB}=0
\ee
provided that $det| h_{\un}{}^a |\neq 0$.

{}From now on we focus on the particular case of $AdS_4$ using the
well-known isomorphism $o(3,2) \sim sp(4|R)$. The algebra $sp(4|R)$
admits the oscillator realization with the generators
\be
\label{al4}
L_{\gal\gb}=\frac1{4i}\{\hat{y}_\gal,
\hat{y}_\gb\}\,,\qquad \,
\bar{L}_{\dot{\gal}\dot{\gb}}
=\frac1{4i}\{\hat{\bar{y}}_{\dot{\gal}},
\hat{\bar{y}}_{\dot{\gb}}\}\,,\qquad \,
P_{\gal\dot{\gb}}=\lambda \frac1{2i}\hat{y}_\gal
\hat{\bar{y}}_{\dot{\gb}}\,
\ee
realized as bilinears in two-component spinor oscillators
satisfying the commutation relations\footnote{One
can equivalently use the Majorana spinor oscillators
$\hat{Y}_\nu$ $(\nu = 1\div 4$) with the commutation relations
$[ \hat{Y}_\mu ,\hat{Y}_\nu ] =2iC_{\mu\nu}\,,$
where $C_{\mu\nu}$ is the charge conjugation matrix.
The language of two-component spinors is however most useful
for the analysis below.}
\be
\label{osc}
[\hat{y}_\alpha , \hat{y}_\beta ] = 2i \epsilon_{\alpha\beta}\,,\qquad
[\hat{\bar{y}}_{\dot{\alpha}} , \hat{\bar{y}}_{\dot{\beta}} ] = 2i
\epsilon_{\dot{\alpha}\dot{\beta}}\,,\qquad
[\hat{y}_\alpha , \hat{\bar{y}}_{\dot{\beta}} ] =0\,
\ee
($\ga ,\gb = 1,2$, $\dga ,\dgb = 1,2 ;\,\bar y_{\dal}=(y_\al)^{+}$; for
 conventions see Appendix).

The $AdS_4$ gravitational fields can now be identified with the 1-form
bilinear in the oscillators
\begin{eqnarray}
\label{spgf}
w_0 = dx^\un w_{0 \un}=\frac{1}{4i}dx^\un
(\omega_{0 \un}{}^{\alpha \beta} \hat{y}_{\alpha}
\hat{y}_{\beta}+\bar \omega_{0 \un}{}^{\dot \alpha \dot \beta}
\hat{\bar{y}}_{\dot \alpha}\hat{\bar{y}}_{\dot\beta}+2\lambda
h_{0 \un}{}^{\alpha \dot \beta} \hat{y}_\alpha \hat{\bar{y}}_{\dot \beta})
\end{eqnarray}
and satisfying the zero-curvature equation
\be
\label{R0}
0=R_0 \equiv d w_0 - w_0 \wedge w_0 \,.
\ee
Here $\omega_{0\un}{}^{\ga\gb} (x)$ and
$\bar{\omega}_{0\un}{}^{\dga\dgb} (x)$
describe Lorentz connection while  $  h_{0\un}{}^{\ga\dgb}(x) $
describes vierbein  in terms of two-component spinors.

The curvature $R =d w - w\wedge w $
admits the expansion analogous to (\ref{spgf}) with the components
\begin {eqnarray}
\label{cur_start}
R_{\alpha \beta}=d\omega_{\alpha \beta}+\omega_\alpha{}^\gamma \wedge
\omega_{\beta \gamma}+\lambda^2 h_{\alpha}{}^{\dot \gamma} \wedge h_{\beta
\dot \gamma}\,, \\
\bar R_{\dot \alpha \dot \beta}= d\bar \omega_{\dot \alpha \dot \beta}+\bar
\omega_{\dot \alpha}{}^{\dot \gamma} \wedge \bar \omega_{\dot \beta \dot
\gamma}+\lambda^2 h^{\gamma}{}_{\dot \alpha} \wedge h_{\gamma \dot \beta}\,,
\\
R_{\alpha \dot \beta}=dh_{\alpha \dot \beta}+\omega_\alpha{}^\gamma \wedge
h_{\gamma \dot \beta}+\bar \omega_{\dot \beta}{}^{\dot \gamma} \wedge
h_{\alpha \dot \gamma}\,.
\label{cur_end}
\end {eqnarray}

A particular solution of (\ref{R0}) can be chosen in the form
\be
\label{thet}
h_{0\un} {}^{\ga\dgb} =-z^{-1} \sigma _\un {}^{\ga\dgb}\,,
\ee
\be
\go_{0\un} {}^{\ga\ga} =-\lambda^2 z^{-1} \sigma _\un
{}^{\ga\dgb}x^\ga{}_\dgb\,,
\ee
\be
\label{thet1}
\bar{\go}_{0\un} {}^{\dgb\dgb}
=-\lambda^2 z^{-1} \sigma _\un {}^{\ga\dgb}x_\ga{}^\dgb\,,
\ee
where
\be
\label{zop}
 x_{\ga \dgb}=\sigma_{\ga \dgb }^a x_a \,, \qquad
x^2=x_a x^a=\half x_{{\ga \dgb}}x^{\ga \dgb} \,,\qquad
z=1+\lambda^2 x^2\,,
\ee
and sigma-matrices are Hermitian
$\bar{\sigma}_\un {}^{\ga\dgb}={\sigma}_\un {}^{\gb\dga}$ with
the normalization
${\sigma}_n {}^{\ga\dgb}{\sigma}_m {}_{\ga\dgb}=2\eta_{nm}$ where
$\eta_{nm} = \{1,-1,-1,-1\}$ is the flat Minkowski metric.
Following to \cite{V2} we use conventions with upper(lower) indices
denoted by the same
letter subject to symmetrization (see also Appendix).

The equations (\ref{thet})-(\ref{thet1}) describe the
vierbein and Lorentz connection
of $AdS_4$ corresponding to the stereographic
coordinates for the hyperboloid realization of $AdS_4$
via embedding into the 5d flat space with  the signature
$(+,-,-,-,+)$. The $AdS_4$ metric tensor resulting from
(\ref{thet}) is
\be
g_{0\um\un}=\half h_{0\um} {}^{\ga\dgb}h_{0\un} {}_{\ga\dgb}=
\frac{\eta_{\um \un}}{{(1+\lambda^2 x^2)}^2}\,.
\ee
Note that this form of the metric can be fixed by the requirement
that it is conformally flat, regular at $x=0$ and
depends on the coordinates
$x^{\ga\dgb}$  via the Lorentz covariant combination $x^2$.
In the flat limit $\lambda \to 0$ this metric describes the
Minkowski space.

In these coordinates,
the boundary of $AdS_4$ is identified with the hypersurface $z=0$.
The ``north pole'' is realized as infinity $x=\infty$.
It becomes a regular point in the inversed coordinates
\be
\label{inver}
x^\un \to x^\un_I = \lambda^{-2} \frac{x^\un}{x^2}
\ee
with the gravitational fields of the form:
\be
\label{thet_in}
h^I_{0\un} {}^{\ga\dgb} =-z^{-1}_I \sigma _\un {}^{\ga\dgb}\,,
\ee
\be
\go^I_{0\un} {}^{\ga\ga} =-\lambda^2 z^{-1}_I \sigma _\un
{}^{\ga\dgb}x_I{}^\ga{}_\dgb\,,
\ee
\be
\label{thet1_in}
\bar{\go}^I_{0\un} {}^{\dgb\dgb}
=-\lambda^2 z^{-1}_I \sigma _\un {}^{\ga\dgb}x_{I\ga}{}^\dgb\,,
\ee
This coordinate system describes the antipodal chart.

\subsection{Star-Product and Free Massless Equations}
\label{Star-Product and Free Massless Equations}

Instead of working in terms of the operators $\hat{y}_\ga$
and $\hat{\bar{y}}_\dga$ it is convenient to use the star-product defined
by the formula
\begin{eqnarray}
\label{star}
(f*g)(y,\bar{y}|x)={(2 \pi)}^{-4} \int d^2 u d^2 \bar{u} d^2 v d^2 \bar{v}
 f(y+u,\bar{y}+\bar{u}|x)g(y+v,\bar{y}+\bar{v}|x) e^{i(u_\al v^\al+\bar
u_{\dal} \bar v^{\dal})} \,.
\end{eqnarray}
It is elementary to see that this star-product is associative
and well-defined for polynomial functions. The normalization is chosen
such that 1 is the unit element of the algebra, i.e. $1*f=f*1 =f$.
The commutation relations for
the generating elements $y_\ga$ and $\bar{y}_\dga$ are
\be
\label{oscstar}
[y_\alpha , y_\beta ]_{*} = 2i \epsilon_{\alpha\beta}\,,\qquad
[\bar{y}_{\dot{\alpha}} , {\bar{y}}_{\dot{\beta}} ]_* = 2i
\epsilon_{\dot{\alpha}\dot{\beta}}\,,\qquad
[{y}_\alpha ,{\bar{y}}_{\dot{\beta}} ]_* =0\,,
\ee
where
$
[a,b]_* = a*b -b*a\,.
$
The formula (\ref{star}) is equivalent to the standard differential Moyal
star-product \cite{MOY} by virtue of the Taylor expansion
\be
f(y,\bar{y})=exp( y^\ga \frac{\partial}{\partial z^\ga} +
\bar{y}^\dga \frac{\partial}{\partial \bar{z}^\dga })
f(z,\bar{z})|_{z=\bar{z}=0}\,.
\ee
It describes the totally symmetric (i.e., Weyl) ordering of the operators
$\hat{y}_\ga ,
\hat{\bar{y}}_\dga $ in terms of symbols of operators and was called
``triangle formula" in \cite{BS}.
The star-product (\ref{star}) acts on the auxiliary spinor variables
$y_\ga , \bar{y}_\dga $ rather than directly on the space-time
coordinates as in the non-commutative Yang-Mills limit of the string
theory \cite{SW}. It was argued however in \cite{rev}
that the dynamical field equations
for higher spin massless fields transform the
star-product nonlocality in $y_\ga , \bar{y}_\dga $ into
the space-time nonlocality of the higher spin interactions.

In terms of the star-product, the equation of the background $AdS_4$ space
has the form:
\be
\label{1}
d w_0 = w_0 *\wedge w_0\,.
\ee

A less trivial fact shown in \cite{V2,Ann} is that free equations
for massless fields of all spins in $AdS_4$ can be cast into the
form
\bee
\label{3}
\ls d w_1 -w_0 *\wedge w_1 - w_1 *\wedge w_0 =
\frac {i}{4} [ h_\alpha{}^{\dot \beta} \wedge h^{\alpha \dot \gamma}
\frac{ \partial }{\partial \bar y^{\dot \beta}} \frac{\partial}{\partial
\bar
y^{\dot \gamma}}C(0,\bar y)+
h^\alpha{}_{\dot \beta} \wedge h^{\gamma \dot \beta }
\frac{ \partial }{\partial y^\alpha} \frac{\partial}{\partial y^\gamma}
C(y,0)]\,,
\end{eqnarray}
\be
\label{2}
d C =w_0 * C- C * \tilde w_0 \, .
\ee

Here $w_1 (y,\bar y |x)$ and $C(y,\bar y|x)$ are functions
of spinor and space-time coordinates and tilde is defined according to
\be
\tilde f(y,\bar y) =
f(y,-\bar y)\,.
\ee

Relativistic fields are identified with the coefficients in the
Taylor expansions in powers of the auxiliary spinor variables
\begin{equation}
\label{gen}
w_1 ({y},{\bar{y}}\mid x)=
\sum_{n,m=0}^{\infty}
\frac{\lambda^{1-|\frac{n-m}{2}|}}{2i\,n!m!}
w_1^{\alpha_1\ldots\alpha_n}{}_,{}^{\dot{\beta}_1
\ldots\dot{\beta}_m}(x)
{y}_{\alpha_1}\ldots {y}_{\alpha_n}{\bar{y}}_{\dot{\beta}_1}\ldots
{\bar{y}}_{\dot{\beta}_m}
\,
\end{equation}
and
\be
\label{Cyby}
C (y,{\bar{y}}|x)=
\sum^\infty_{n,m=0}
\frac{\lambda^{2-\frac{n+m}{2}}}{m!n!} C^{\gal_1\ldots\gal_n}{}_{,}{}^
{{\dga}_1 \ldots {\dga}_m}(x)
y_{\gal_1}\ldots y_{\gal_n}\,
{\bar{y}}_{\dga_1}\ldots {\bar{y}}_{\dga_m}\,,
\ee

Inserting (\ref{Cyby}) into (\ref{2}), one
arrives
at  the following infinite chain of equations
\begin{equation}
\label{DCC}
D^LC_{\alpha (m),\,\dot{\beta}
(n)}= i h^{\gamma\dot{\delta}}C_{\alpha (m)\gamma,\, \dot{\beta} (n)
\dot{\delta}} -i\lambda^{2} nm h_{\alpha\dot{\beta}} C_{\alpha
(m-1),\,\dot{\beta}
(n-1)}\,,
\end{equation}
where $D^L$ is the Lorentz-covariant differential
\begin{equation} D^L
A_{\alpha\dot{\beta}}=dA_{\alpha\dot{\beta}}+\omega_\alpha{}^\gamma \wedge
A_{\gamma\dot{\beta}}+\bar{\omega}_{\dot{\beta}}{}^{\dot{\delta}} \wedge
A_{\alpha\dot{\delta}}\,.
\end{equation}
The system (\ref{DCC}) decomposes into a set of independent
subsystems with $n-m$ fixed. It turns out \cite{Ann}
that the subsystem
with $|n-m| = 2s$ describes a massless field of spin $s$
(note that the fields
$C_{\alpha (m),\,\dot{\beta}(n)}$
and
$C_{\beta (n),\,\dot{\ga} (m)}$ are
complex conjugated).

Analogously, by the substitution of (\ref{gen}),
the equation (\ref{3}) amounts to
\bee
\label{R1}
\ls R_{1\,\alpha (n)\,,\dgb (m)} \ls & \equiv D^L w_{1\;\alpha (n)\dgb (m)}
+n\gamma( n,m,\lambda ) h_{\alpha}{}^{\dot{\delta}}\wedge
w_{1\;\alpha (n-1)\,,\dgb (m) \dot{\delta}}
+m \gamma( m,n,\lambda )  h^{\gga}{}_{\dgb}\wedge
w_{1\;\gga \alpha (n)\,,\dgb (m-1) } \nonumber \\
&=\delta (m)  h^{\gamma \dot{\delta}}
\wedge h^{\gamma}{}_{\dot{\delta}} C_{\alpha (n) \gamma (2)} +
\delta(n)   h^{\eta\dot{\delta}}\wedge h_\eta{}^{\dot{\delta}}
\bar{C}_{\dot{\beta} (m) \dot{\delta}(2)}\,,
\eee
where
\be
\gamma( n,m,\lambda ) =\theta (m-n)+\lambda^{2} \theta(n-m-2)+
\lambda\delta(n-m-1).
\ee

A spin $s\geq 1$ dynamical massless field is identified with the 1-form
(potential)
\be
\label{physi}
w_{\ga (n),\dgb (n)}\phantom{MMMMMMMM}  \qquad n = (s-1)\,,
\qquad s\geq 1 \quad \mbox{integer},
\ee
\be
\label{physo}
w_{\ga (n),\dgb (m)}
  \qquad n+m = 2(s-1)\,, \quad |n-m|
=1 \qquad s\geq 3/2 \quad \mbox{half-integer}.
\ee

The matter fields are described by the 0-forms
\be
\label{phys0} C_{\ga (0)\,,\dga (0)}\qquad
\qquad\qquad\qquad\qquad\qquad\qquad
s=0\,,
\ee
\be
\label{phys1/2}
C_{\ga (1)\,,\dga (0)}     \oplus C_{\ga (0)\,,\dga (1)} \qquad
\qquad\qquad\qquad\qquad s=1/2\,.
\ee

All other components of the expansions (\ref{gen})
and (\ref{Cyby}) express in terms of the derivatives
of physical fields by virtue of the equations (\ref{R1}), (\ref{DCC}).
For example, for 0-forms $C$ one has

\be
\label{Cn>m}
\!\!C_{\ga (n)\,,\dgb (m) }\!=\!\frac{1}{(2i)^{\half (n+m -2s)}}
h^{\un_1}_{\ga\dgb}  D^L_{\un_1}
\!\ldots h^{\un_{\half (n+m -2s)}}_{\ga\dgb}\!
D^L_{\un_{\half (n+m -2s)}}\! C_{\ga (2s)}\,\quad n \geq m ,
\ee
or
\be
\label{Cn<m}
\!\!C_{\ga (n)\,,\dgb (m) }\!=\!\frac{1}{(2i)^{\half (n+m -2s)}}
h^{\un_1}_{\ga\dgb} D^L_{\un_1}
\!\ldots  h^{\un_{\half (n+m -2s)}}_{\ga\dgb} \!
D^L_{\un_{\half (n+m -2s)}} C_{\dgb (2s)}\,\quad n \leq m .
\ee

Analogous formula holds for the 1-forms $w$

\be
w_{\ga (n)\,,\dgb (m)} \sim \left (
\frac{\partial}{\partial x}
\right )^{[\frac{|n-m|}{2}]} w^{phys} +w^{gauge}\,,
\ee
where $w^{phys}$ denotes the appropriate field from the list
(\ref{physi}), (\ref{physo}) while $w^{gauge}$
is a pure gauge part.

As shown in \cite{V2,Ann} the content of the equations (\ref{3}) and
(\ref{2}) is equivalent to the relations (\ref{R1}), (\ref{DCC}) and usual
dynamical equations for the physical massless fields in $AdS_4$.
The fields
\be
w^{\alpha(n),\,\dot \beta(m)}
\quad \mbox{with} \quad
 n+m=2(s-1)
\ee
and
\be
C_{\alpha (m),\,\dot{\beta}(n)}
\quad \mbox{with} \quad |n-m| = 2s
\ee
are associated with the massless field of spin $s$ along with
 all its on-mass-shell nontrivial derivatives.
For spin $s\geq 1$, the 0 forms $C$ describe gauge invariant
field strengths generalizing the spin 1 Maxwell field strength and
spin 2 Weyl tensor to an arbitrary spin.  For example, in the spin 2
case (\ref{gen}) is equivalent to the linearized Einstein
equations because it just tells us that torsion is zero and all
the components
of the linearized Riemann tensor are zero except for those which are
described by the Weyl tensor $C_{\ga_1 \ldots \ga_4}$,
$C_{\dga_1 \ldots \dga_4}$.

The aim of this paper is to explore the fact
that once the dynamical equations
are reformulated in the form (\ref{3}) and (\ref{2}) one can write
down their generic solution explicitly in terms of the
star-product provided that the $AdS_4$ vacuum equations
(\ref{1})-(\ref{2}) are solved in the pure gauge form
\be
\label{1s}
w_0=-g^{-1} * dg\,,
\ee
where $g(y,\bar{y}|x)$ is some invertible element of the
star-product algebra, $ g*g^{-1}=g^{-1}*g=1$ and
$d=dx^\un \frac{\partial}{\partial x^\un}$
is the space-time differential. A form of the appropriate
$g(y,\bar{y}|x)$ is found in the section
\ref{Gauge Function}. The covariant constancy equation (\ref{2}) has
the generic solution of the form
\be
\label{2s}
C(x)=g^{-1}(x)*C_0*\tilde g(x)\,,
\ee
where $C_0 (y,\bar{y}) $ is an arbitrary $x-$independent
element of the star-product algebra. For exponential
functions $C_0$ we reproduce in the section
\ref{ Plane Waves } a set of particular solutions that
 in the flat limit tend to the plane wave solutions for
massless fields of an arbitrary spin.
A less trivial fact shown in the section
\ref{Higher spin potentials}
is that it is also possible to write down a generic solution of
the equation (\ref{3})
despite  it does not have a zero-curvature form.

Let us emphasize that this way of solving
dynamical equations reduces to evaluation of some elementary Gaussian
integrals originating from the star-product (\ref{star})
(i.e. there is no need to solve any differential equations). The
equation (\ref{2s}) can be interpreted as a  covariantized Taylor
expansion. Indeed, if $g(x_0 )=1$ for some $x_0$ then
$C_0 (y,\bar{y})=C(y,\bar{y}|x_0 )$.
In accordance with (\ref{Cyby}),
(\ref{Cn>m}) and (\ref{Cn<m}),  $C_0 (y,\bar{y})$ therefore
describes all on-mass-shell
nontrivial derivatives of the physical field at the point $x_0$.

\section {Gauge Function}
\label{Gauge Function}

In this section we find the function $g(y,\bar{y}|x)$ that reproduces
the $AdS_4$ gravitational fields (\ref{thet})-(\ref{thet1}) in the
star-product algebra version of the ansatz (\ref{spgf})
\begin{eqnarray}
\label{basic_form}
w_{0 \un}=\frac{1}{4i}(\omega_{0 \un}{}^{\alpha \beta} y_{\alpha}
y_{\beta}+\bar
\omega_{0 \un}{}^{\dot \alpha \dot \beta}\bar y_{\dot \alpha}\bar y_{\dot
\beta}+2\lambda
h_{0 \un}{}^{\alpha \dot \beta} y_\alpha \bar y_{\dot \beta})
\end{eqnarray}
by virtue of the pure gauge representation (\ref{1s}).
Note that the bilinear ansatz is consistent, because the bilinears form
a closed $sp(4)$ subalgebra with respect to
commutators\footnote{It is a simple exercise with the star-product to
check that this ansatz with the fields (\ref{thet})-(\ref{thet1})
satisfies (\ref{1})}. We therefore have
to find such a function $g$ that the resulting connection
(\ref{1s}) is bilinear in the auxiliary oscillators $y$ and $\bar{y}$.
Note that if $w_0$ would contain some higher-order polynomials in
oscillators, this would imply that some higher spin fields acquire
nonvanishing vacuum expectation values thus making the physical
interpretation of the corresponding solutions less straightforward.

Let us look  for
the solution for $g$ in the Lorentz-covariant form
\be
g(y,\bar y|x)=e^{if(x^{2})x^{\alpha \dot \alpha}y_{\alpha}\bar y_{\dot
\alpha}+r(x^{2})}
\ee
with some functions $f(x^2 )$ and $r(x^2 )$.
One readily shows that the inverse element $g^{-1}$ is
\be
g^{-1}(y,\bar y|x)=(1+f^2 x^2)^2 e^{-if(x^{2})x^{\alpha \dot
\alpha}y_{\alpha}\bar y_{\dot \alpha}-r(x^{2})}\,.
\ee
Direct computation using the star-product (\ref{star}) leads
by virtue of evaluation of elementary Gaussian integrals to
the following result
\bee
\label{first_result}
g^{-1} * dg = 2(m_1 +m_2 x^{\al \dbt} y_\al \bar{y}_{\dbt}
)x_n dx^n+ m_3 y_\al \bar{y}_{\dbt} dx^{\al \dbt}
+m_4 ( x^{\gamma}{}_{\dbt} y_\al y_\gamma +
x_{\al}{}^{\dot\gamma} \bar y_{\dbt} \bar y_{\dot \gamma})
dx^{\al \dbt} \,,
\eee
where
$$m_1 = r^\prime - q^{-1}(2ff^\prime x^2+f^2), $$
$$m_2 = i q^{-1}  f^\prime +i q^{-2} f^3\,,$$
$$m_3 = iq^{-2}f(1-f^2x^2)\,,$$
$$m_4 = iq^{-2}f^2\,,$$
and we use notation
$$ q=1+f^2 x^2  $$
and
$$
 h^\prime (x^2)= \frac{\partial h (x^2 )}{\partial x^2}\,.
$$

In order to reproduce $AdS_4$ vierbein and connection
(\ref{thet})-(\ref{thet1}) we have to set $m_1=m_2 =0$.
The condition $m_2 =0$ is equivalent to
\be
\label{dif_eq}
f^3+ f^\prime+f^\prime f^2 x^2=0.
\ee

Its general solution is
\begin{eqnarray}
\label{fsol}
f_{\pm}= \lambda (\frac{1}{1 \pm \sqrt z})\,,
\end{eqnarray}
where $\lambda$ is an arbitrary
integration constant and $z$ is defined in
(\ref{zop}). The solution $f_+$ is regular at $x\to 0$.
The solution $f_-$ is singular at $x\to 0$. In fact, the two
gauge functions $g$ corresponding to these
solutions are related to each other by the inversion (\ref{inver})
away from the north ($x=\infty $) and south ($x=0$) poles
thus solving the problem for the two stereographic charts.
In the rest of the paper we will for definiteness consider the
solution regular at $x=0$ with $f=f_+$.

Solving the condition $m_1 =0$ for $f_{+}$, one finds up  to the
integration constant that does not affect $w_0$, that
\be
\label{rsol}
r=ln (\frac{2\sqrt{z}}{1+\sqrt{z}}) .
\ee
As a result $g$ takes the form
\begin {eqnarray}
\label{gz}
g(y,\bar{y}| x) = 2\frac{\sqrt{z}}{1+\sqrt{z}}
\exp[\frac{i\lambda}{1+\sqrt{z}}x^{\al \dal}y_{\al} \bar y_{\dal} ]
\end {eqnarray}
with the inverse
\be
\label{g-1}
g^{-1}(y,\bar{y}| x) = \tilde{g}(y,\bar{y}| x) =
2\frac{\sqrt{z}}{1+\sqrt{z}}
\exp[\frac{-i\lambda}{1+\sqrt{z}}x^{\ga\dgb}y_\ga \bar{y}_\dgb ]\,.
\ee
 Inserting (\ref{fsol}) and (\ref{rsol}) into
the expressions for $m_3$ and $m_4$ we find that the corresponding
Lorentz connection and vierbein  reproduce (\ref{thet})-(\ref{thet1}).
The solution (\ref{thet_in})-(\ref{thet1_in}) corresponds to $f_{-}$.

\section { Plane Waves }
\label{ Plane Waves }

Having found the gauge function $g$ one solves
the free field equation (\ref{2}) for matter fields and
Weyl tensors of an arbitrary spin in the form (\ref{2s})
equivalent to
\bee
\label{C}
C(y,\bar y|x) = (2 \pi)^{-4} \int_{-\infty}^{\infty} &{}&\ls
d^2ud^2vd^2\bar u d^2\bar v
       \exp i(u_\ga v^\ga+\bar u_\dga\bar v^\dga) \nn\\
&{}&\ls\ls\times  g^{-1}(y+u,\bar y+\bar u)
C_0(y+u+v,\bar y + \bar u + \bar v)
\tilde{g}(y+v,\bar y+\bar v)\,.
\eee
As emphasized in the section
\ref{Star-Product and Free Massless Equations},
this formula reproduces the covariantized
Taylor expansion with $C_0$ identified with all on-mass-shell nontrivial
derivatives of $C(x)$ at $x=x_0$ with $g(x_0 )=I$. Note that
$x_0=0$ is the ``south pole" for the solution (\ref{gz}).

Let us now choose $C_0$ in the form
\begin{eqnarray}
\label{cdef}
C_0=\lambda^2 c_0 \exp i\lambda^{-\half}
(y^\alpha \eta_\alpha + \bar y^{\dot \alpha} \bar \eta_{\dot
\alpha})\,,
\end{eqnarray}
where $c_0$ is an arbitrary constant and
$\eta_\ga$ and $\bar{\eta}_\dga$ are complex conjugated
constant spinor parameters. Substitution to (\ref{C}) leads
 to the following result upon evaluation of the elementary
Gaussian integrals
\be
\label{solut}
C(y,\bar y|x) = c_0 \lambda^2 z \exp i\left
[ -(\lambda y_\ga \by_\dgb+\eta_\ga\bar{\eta}_\dgb )x^{\ga\dgb}
+\lambda^{-\half}
\sqrt{z}(y^\ga \eta_\ga +\by^\dga \bar{\eta}_\dga )\right ]\,.
\ee

Taking into account
the identification of the particular components of the expansion
(\ref{Cyby})  explained in the section
\ref{Star-Product and Free Massless Equations}
one identifies
the matter fields and higher spin Weyl tensors with
\be
\label{rC}
C_{\ga_1 \ldots \ga_n} (x) = \lambda^{\frac{n}{2}-2}
\frac{\partial}{\partial y^{\ga_1}}
\ldots \frac{\partial}{\partial y^{\ga_n}}
C(y,\by |x) |_{y=\by =0}
\ee
and
\be
\label{cC}
\bar C_{\dot \ga_1 \ldots \dot \ga_n} (x) =
\lambda^{\frac{n}{2}-2 }
\frac{\partial}{\partial \bar
y^{\dot \ga_1}}
\ldots \frac{\partial}{\partial \bar y^{\dot \ga_n}}
C(y,\by |x) |_{y=\by =0}\,.
\ee

{}From (\ref{solut}) one gets
\be
\label{1eq}
C_{\ga \ldots \ga_{2s}} (x) =  c_0 z^{s+1}
\eta_{\ga_1} \ldots \eta_{\ga_{2s}}
\exp i k_{\gamma\dgb} x^{\gamma\dgb} \,
\ee
and
\be
\label{2eq}
\bar C_{\dot \ga \ldots \dot \ga_{2s}} (x) = c_0 z^{s+1}
\bar \eta_{\dot \ga_1} \ldots \bar \eta_{\dot \ga_{2s}}
\exp i k_{\gamma\dgb} x^{\gamma\dgb} \,,
\ee
where
\be
\label{twist}
k_{\ga\dgb} =-  \eta_\ga \bar{\eta}_\dgb\,.
\ee
The expression (\ref{twist})
is  the standard twistor representation \cite{PR} for an
arbitrary light-like vector $k_{\ga\dgb} $ in terms of spinors.

By construction, the formulae (\ref{1eq}) and (\ref{2eq})
describe solutions of the free equations of motion of massless
fields of arbitrary spin in $AdS_4$.
In the flat space  $\lambda \to 0$  (i.e., $z\to 1$)
these formulae describe  usual flat space plane waves. We therefore
interpret the solution (\ref{solut}) as describing $AdS_4$ ''plane waves".
Let us note that, as expected, these solutions tend to zero
at the boundary of $AdS_4$  $z=0$.

\section {Higher Spin Potentials}
\label {Higher spin potentials}

Once the 0-forms $C$ are found one can in principle solve the
equation (\ref{3}) for the gauge potentials $w_1$
modulo gauge transformation
\be
\delta w_1 = d\epsilon -w_0 * \epsilon + \epsilon * w_0 .
\ee
where $w_0$ is the background $AdS_4$ gauge (gravitational) field
(\ref{basic_form}) and
$\epsilon (y,\bar{y}|x) $ is an arbitrary gauge parameter.

This problem is analogous to the reconstruction of the electromagnetic
potential via the field strength or the metric tensor
via the  Riemann tensor.
Since the equation (\ref{3}) is formally consistent it admits some
solution. Remarkably, this solution can also be found explicitly
for the general gauge function $g (y,\bar{y}|x)$ and $C_0 (y,\bar{y})$.
The systematic derivation can be obtained with the help of a
more sophisticated technics developed in \cite{PV,rev} for the analysis
of the nonlinear problem. Here we announce the final result
\bee
\label{pot}
w_1(y,\bar y|x) & = & \frac{1}{32\pi^4} (\omega_{0 }{} ^{\al \bt}
\frac{\partial}{\partial
y^\bt} + \lambda h_{0 }{}^{\al \dbt}
\frac{\partial}{\partial \bar y^{\dbt} })
\int_0^1 s ds \int_{-\infty}^{\infty} d^2u d^2v d^2 \bar u d^2 \bar v
[ (y_\al+2v_\al) \nonumber \\
&&\times g^{-1} ((1-s)y+u,\bar y+\bar u |x)
C_0((1-s)y+v+u,\bar y + \bar u + \bar v)
\nonumber \\
&&\times g((1-s)y+(1-2s)v,\bar y +\bar v|x) \exp i(uv+ \bar u \bar v )]+c.c.\,,
\eee
leaving the details of its derivation to a more detailed future
publication \cite{BV}.

Note that this expression admits no simple interpretation
in terms of the star-product (\ref{star}). This happens because it
is derived \cite{BV} as some projection from a larger star-product
algebra. However, one can  check directly  that (\ref{pot}) solves
 (\ref{3}). To this end one  inserts this formula into
the left hand side of (\ref{3}) using the equations (\ref{2}).
Then one observes that the resulting terms in the integrand sum up
to a total derivative in the integration variable $s$. The contribution
at $s=0$ vanishes because of the factor of $s$ in the integration measure,
while the contribution at $s=1$ reproduces the right hand side
of (\ref{3}). Note that only the terms independent of the Lorentz connection
 contribute. The computation sketched above is however tedious enough.

Insertion of the expression (\ref{cdef}) into (\ref{pot}) leads to
the following result
\bee
\label{potp}
{}&& w_1=\half z c_0 (\omega^{\alpha
\beta}\frac{\partial}{\partial y^{\beta}}+\lambda h^{\alpha \dot
\gamma}\frac{\partial}{\partial \bar y^{\dot \gamma}})
 \int_0^1ds \frac{s}{(s+(1-s)\sqrt{z})^3} \nn\\
&{}&\times (y_\ga +\lambda^{-1/2} (1+\sqrt{z}) \eta_\ga
+\lambda^{1/2}x_{\ga\dgb}\bar{\eta}^{\dgb} -\lambda \bar{y}^\dgb
x_{\ga\dgb} )\nn\\
{}&&\times \exp i \frac{\lambda^{-1/2}}{s+(1-s)\sqrt{z}}
 \Big [ (1-s)(\eta_\ga y^\ga  + \bar\eta_{\dal} \bar y^{\dal}
+\lambda x^{\ga\dgb} (y_\ga \bar{\eta}_\dgb +\eta_\ga\bar{y}_\dgb ))\nn\\
&{}&\phantom{\times \exp i \frac{\lambda^{-1/2}}{s+(1-s)\sqrt{z}}}
+s (\sqrt{z}\bar\eta_{\dal} \bar y^{\dal}
-\lambda^{1/2}x^{\ga\dgb} \eta_\ga \bar{\eta}_\dgb )\Big ]
+c.c.
\eee
Let us note that the quantity $s+(1-s)\sqrt{z}$ is strictly positive in
the region
$z>0$
for any $0\leq s\leq 1$. Therefore, the right hand side of the
formula (\ref{potp}) is some entire function in the auxiliary
spinor variables $y$ and $\bar{y}$. This means that it
gives rise to the
well-defined gauge potentials associated with the
coefficients of the expansion (\ref{gen})
 everywhere in the north pole chart. The south pole chart
can be analyzed analogously.
In the limit $z\to0$
the expression (\ref{potp}) acquires singularity at $s=0$.


\section {Conclusion}
It is demonstrated that the general formalism developed
originally for the formulation of the interacting higher spin gauge
theories allows one to write down explicit solutions of the field equations
in terms of the star-product algebras in auxiliary spinor variables.
The true reason for this is that the formalism of
star-product algebras makes  infinite-dimensional higher spin
symmetries explicit thus allowing one to formulate the field
equations as certain
covariant constancy conditions. For simplicity, in this paper we have
focused on the most symmetric case with gravitational
fields possessing explicit Lorentz symmetry. The developed approach
can be applied in  other coordinate systems.
We believe that it will have a wide area of applicability and can be
extended to dynamical systems
in various dimensions, black hole and brane backgrounds as well as
to the superspace.

\section*{Acknowledgments}

The authors are grateful
to R.Metsaev, S.Konstein and D.Sorokin for useful conversations.
This research was supported in part by
INTAS, Grant No.96-0308 and by the RFBR Grant No.99-02-16207.

\section*{Appendix. Notation}
Following to \cite{V2},
we use conventions with upper(lower) indices denoted by the same
letter subject to symmetrization
prior possible contractions with lower(upper) indices denoted by the
same letter. With these conventions only a number of indices within
any symmetrized group is important. This number is often indicated
in brackets. A maximal possible number
of lower and upper indices denoted by the same letter
is supposed to be contracted.

We choose mostly  minus flat  Minkowsi metric $\eta^{nm} = \{1,-1,-1,-1\}$.
$\partial_\un =\frac{\partial}{\partial x^\un}$.

Two-component spinor indices are raised and lowered according to:

$$A_\alpha=\e_{\beta \ga}A^\beta \,\qquad
A^\alpha=\e^{\ga \beta} A_\beta \,,$$
where
$e_{1 2}=e^{1 2}=1$.
Sigma matrices are expressed in terms of Pauli matrices according to
$$\sigma^m_{\alpha \dot \beta}=(I,\sigma^i)$$
$$\sigma^n_{\alpha \dot \beta}\sigma^{m \alpha \dot \beta}=2\eta^{n m}\,.$$
Also the following conventions are used
$$ x_{\ga \dgb}=\sigma_{\ga \dgb }^a x_a \,, \qquad
x^2=x_a x^a=\half x_{{\ga \dgb}}x^{\ga \dgb}  .$$

$$\delta(n)=
\left\{
\begin {array}{ll}
1 \, , \, n=0\\
0 \, , \, n  \ne 0\\
\end {array}
\right. \,,
$$

$$\theta(n)=
\left\{
\begin {array}{ll}
1 \, , \, n \ge 0\\
0 \, , \, n  < 0\\
\end {array}
\right . \,.
$$

\end{document}